\documentclass{elsarticle}
\usepackage{xcolor}
\usepackage{bm}
\usepackage{graphicx}
\usepackage{amsfonts}
\usepackage{amsmath}
\usepackage{subfigure}
\usepackage{float}
\usepackage{lineno,hyperref}

\usepackage{graphicx}

\usepackage{subfigure}
\usepackage[margin=1in]{geometry}
\usepackage{amssymb}
\usepackage{amsmath}

\usepackage{float}
\usepackage{url}
\usepackage{xspace}
\usepackage{hyperref}
\usepackage{xcolor}

\usepackage{amsmath}
\usepackage{tikz}
\usetikzlibrary{automata,arrows,positioning,calc}

\begin{document}

\begin{frontmatter}

\title{Towards the Modeling of Behavioral Trajectories of Users in Online Social Media}

\author{Alessandro Bessi}
\corref{mycorrespondingauthor}
\cortext[mycorrespondingauthor]{Corresponding author}
\ead{bessi@isi.edu}

\address[mymainaddress]{University of Southern California, Information Science Institute, Marina del Rey, CA (USA)}

\begin{abstract}
In this paper, we introduce a methodology that allows to model behavioral trajectories of users in online social media. First, we illustrate how to leverage the probabilistic framework provided by Hidden Markov Models (HMMs) to represent users by embedding the temporal sequences of actions they performed online. We then derive a model-based distance between trained HMMs, and we use spectral clustering to find homogeneous clusters of users showing similar behavioral trajectories. To provide platform-agnostic results, we apply the proposed approach to two different online social media --- i.e. Facebook and YouTube. We conclude discussing merits and limitations of our approach as well as future and promising research directions.
\end{abstract}

\end{frontmatter}

\section*{Introduction}
Over the last decade, the rise of online social media has caused a huge shift in the way people find information, interpret facts, and shape their opinions. Facebook news feeds, Twitter timelines, and blogs are replacing morning newspapers and nightly news. Nowadays everyone can produce and consume information without any filter or restriction.  

Such a disintermediated environment has proved to be a fiasco for the public's understanding of current affairs: clickbait news that pander to readers' worst instincts are proliferating on blogs \cite{blom2015click}; conspiracy theories that simplify causation and reduce the complexity of reality are spreading more than stories that are balanced and thoroughly reported \cite{del2016spreading,menczer2016spread}; Facebook is flooded by fake news fabricated by fringe websites \cite{bessi2015science, bessi2015trend, bessi2016statistical,shao2016hoaxy}; Twitter is swamped by bots \cite{ferrara2014rise} --- algorithmically driven entities that on the surface appear as legitimate users --- distorting the political debate \cite{bessi2016social}; the emergence of virtual echo chambers --- non-interacting polarized communities centered on different narratives wherein enclaves of like-minded people reinforce their preexisting beliefs \cite{bessi2016personality, zollo2015emotional,bessi2016homophily,del2016anatomy,del2016echo} --- is reducing viewpoint diversity and flattening debates \cite{quattrociocchi2016echo, dewey2012public, habermas2015between, flaxman2016filter,starnini2016emergence,bright2016explaining,galam2016modeling}.

What is happening in online social media is worsening the political polarization, jeopardizing the quality of democratic discourse, influencing policy preferences, and encouraging behaviors strongly divergent from recommended practices. For these reasons, a better understanding of the behavioral, cognitive, and psychological processes underlying the observed dynamics is a matter that Science has to address.

In this work, we propose a methodology that leverages Hidden Markov Models \cite{eddy1996hidden,blunsom2004hidden,rabiner1986introduction} to represent behavioral trajectories of users in online social media. A Hidden Markov Model (HMM) is a probabilistic model in which the system being modeled is assumed to be a Markov process with unobserved (\emph{hidden}) states.  HMMs extend the framework provided by Markov chains in order to model systems in which the states (or events) we are interested in are not directly observable. HMMs are traditionally known for their application in temporal pattern recognition such as speech, gesture recognition, and bioinformatics \cite{krogh2001predicting, bahl1986maximum, yamato1992recognizing}. Recently, the application of HMMs has been successfully extended to computational social science --- e.g. in \cite{dedeo2015conflict} the author provides novel evidence for the existence of an epoch-like structure of conflict and cooperation on Wikipedia, distinguished by behavioral motifs.

The fundamental idea behind the approach that we are introducing is the following. In social network analysis, we can observe actions performed by users --- e.g. likes, comments, shares, retweets, etc. ---, but the worldviews, inclinations, and orientations driving those actions remain hidden. It follows that Hidden Markov Models --- wherein the hidden states are supposed to cause observables outputs --- might provide an appropriate and convenient probabilistic framework for the modeling of behavioral trajectories of users in online social media. In this paper, we show that HMMs can embed time series of different length representing the comments left by users supporting conflicting narratives. For the sake of generalization and to provide platform-agnostic results, we apply our methodology to two different online social media: Facebook and YouTube.

Our results show that Hidden Markov Models are able to model behavioral trajectories of users by embedding their visible actions in online social media. Besides the soundness of the intuition and the straightforward idea motivating the use of HMMs in this context, the main strength of our approach is that it allows to compare users that performed a different number of actions --- i.e. users that are represented by time series of different length. Indeed, we can compare users by using a \emph{model-based} distance between their HMMs, and then apply spectral clustering to discover homogeneous clusters of users showing similar trajectories.

\section*{Materials and methods}
\subsection*{Data Collection}
To test the proposed methodology, we rely on a dataset already used in \cite{bessi2016users}. In particular, we use data available for two random samples of $1.2K$ users that left at least $100$ comments either on Facebook posts or YouTube videos supporting different and conflicting narratives. Indeed, both the posts and videos considered have been published by pages and channels disseminating either Conspiracy or Science news. 

The first category (Conspiracy) includes pages and channels diffusing alternative and controversial information, usually lacking supporting evidence and most often contradictory of the official news. The second category (Science) includes scientific institutions and scientific press having the main mission of diffusing scientific knowledge. Such a space of investigation is defined with the same approach as in \cite{del2016spreading}, with the support of different Facebook groups very active in monitoring the conspiracy narratives. Both pages and channels were accurately selected and verified according to their self description.

The data collection started with the download of all the Facebook posts --- and their respective users' interactions --- published by 419 US Facebook pages supporting either Science or Conspiracy. Then, we collected metadata related to YouTube videos linked by such posts, as well as the associated users' interactions.

The entire data collection process has been carried out exclusively through the Facebook Graph API and the YouTube Data API, which are both publicly available, and for the analysis we used only public available data (users with privacy restrictions are not included in the dataset). The pages from which we download data are public Facebook and YouTube entities. User content contributing to such entities is also public unless the user’s privacy settings specify otherwise and in that case it is not available to us. We abided by the terms, conditions, and privacy policies of the websites (Facebook and Youtube).

\subsection*{Hidden Markov Models}
A Hidden Markov Model (HMM) is a probabilistic model in which the system being modelled is assumed to be a Markov process with unobserved (\emph{hidden}) states.  HMMs extend the framework provided by Markov chains in order to model systems in which the states (or events) we are interested in are not directly observable. The basic structure of a HMM consists of a set of hidden states, each of which produces an observable output (\emph{observation}). A first-order HMM instantiates two simplifying assumptions:

\begin{enumerate}
	\item The probability of a particular state depends only on the previous state:
$$ \mathbf{Pr}(x_{t}| x_{1},\dots,x_{t-1}) = \mathbf{Pr}(x_{t}| x_{t-1}).$$
   \item The probability of an observation $o_{t}$ depends only on the state that produced the observation --- not on any other states or any other observations --- that is:
$$ \mathbf{Pr}(o_{t}| x_{1},\dots,x_{t},\dots,x_{T}, o_{1},\dots,o_{t},\dots,o_{T}) = \mathbf{Pr}(o_{t}| x_{t}).$$
\end{enumerate}

Figure \ref{fig:fig0} provides a graphical representation of a first-order HMM.

\begin{center}
	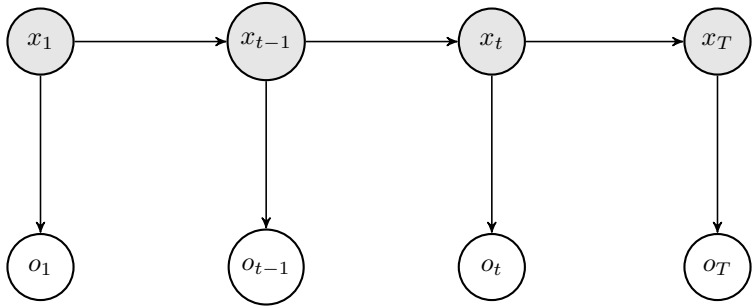
\begin{figure}[ht]

		\centering
		\begin{tikzpicture}[->, >=stealth', auto, semithick, node distance=3cm]
		\tikzstyle{every state}=[fill=white,draw=black,thick,text=black,scale=1]
		\node[state, fill=black!10]    (A)                     {$x_{1}$};
		\node[state, fill=black!10]    (B)[right of=A]   {$x_{t-1}$};
		\node[state, fill=black!10]    (C)[right of=B]   {$x_{t}$};
		\node[state, fill=black!10]    (D)[right of=C]   {$x_{T}$};
		
		\node[state]    (E)[below of=A]                     {$o_{1}$};
		\node[state]    (F)[right of=E]   {$o_{t-1}$};
		\node[state]    (G)[right of=F]   {$o_{t}$};
		\node[state]    (H)[right of=G]   {$o_{T}$};
		
		\path
		(A)
		edge (B)
		edge (E)
		
		(B)
		edge (C)
		edge (F)
		
		(C)
		edge (D)
		edge (G)
		
		(D)
		edge (H)
		;
		
		\end{tikzpicture}
	\caption{\textbf{Graphical representation of a first-order HHM.} Gray shaded nodes represent the hidden states of the system. The probability of a particular state of the system at time $t$ depends only on the state of the system at time $t-1$. White nodes represent the observations produced by the hidden states of the system. The probability of an observation at time $t$ depends only on the (hidden) state of the system at time $t$.} 
	\label{fig:fig0}
	\end{figure}
\end{center}

Formally, a first-order HMM is defined by:

\begin{enumerate}
	\item A set of hidden states $X = \{ X_{1},\dots,X_{|X|} \}$.
	\item A set of visible states (\emph{observations}) $O = \{ O_{1},\dots,O_{|O|} \}$.
	\item A state transition probability matrix $\mathbf{A} = \{ a_{ij} \}$, where $a_{ij} = \mathbf{Pr}(x_{t} = X_{j} | x_{t-1} = X_{i} )$, $1 	\leq i,j \leq |X|$.
	\item An observation probability matrix  $\mathbf{B} = \{ b_{ki} \}$, where $b_{ki} = \mathbf{Pr}(o_{k} | X_{i} )$, $1 	\leq k \leq |O|$, $1 	\leq i \leq |X|$.
\end{enumerate}

\section*{Results}
\subsection*{Modeling behavioral trajectories of users using HMMs}
In this work, we show that is possible to use first-order discrete HMMs to model the behavioral trajectories of users in online social media. The intuition supporting this approach is the following. In social network analysis, we cannot observe the actual orientations of users towards a specific kind of content. Indeed, the only things that we can observe are temporally ordered sequences of actions that users perform --- e.g. tweets, comments, likes, shares, etc. In this context, HMMs provide a convenient as well as intuitive probabilistic framework to model the unobserved (\emph{hidden states}) orientation of users by leveraging their observed (\emph{visible states}) actions in online social media.

Here we illustrate how to use the proposed methodology to model behavioral trajectories of users consuming contents supporting conflicting narratives --- i.e. Science and Conspiracy (see Data Collection for additional information). For the sake of generalization as well as to provide platform-agnostic results, we apply our approach to different online social media --- i.e. Facebook and YouTube.

For both Facebook and YouTube we focus on a random sample of $1.2K$ users with at least $100$ comments. Both samples are composed as follows. A first batch of $400$ users with more than $95\%$ of their comments on posts (videos) supporting Science; a second batch of $400$ users with more than $95\%$ of their comments on posts (videos) supporting Conspiracy; a third batch of $400$ users with no more than $95\%$ of their comments on posts (videos) supporting either Science or Conspiracy. The users in the first and the second batches are considered polarized towards either Science ($P_{S}$) or Conspiracy ($P_{C}$), whereas the users in the third batch are not polarized ($NP$). We choose to consider three balanced batches of users with different orientations to obtain nice and interpretable results. Still, our approach does not need balanced data. Similarly, the choice of considering users with at least $100$ comments is arbitrary, since HMMs can be fitted with sequences of shorter length --- even if the longer the sequence the better the estimation of the HMMs parameters.

Since Facebook and YouTube data share the same structure, hereafter we make no distinction between the two and we refer generically to users and their behavioral trajectories while introducing our proposed methodology.

For each user we instantiate a HMM, $\lambda$, with three hidden states and two visible states. The three hidden states represent the orientation of the user: polarized towards Science ($\mathcal{S}$), uncertain ($\mathcal{U}$), and polarized towards Conspiracy ($\mathcal{C}$). The two visible states (observations) are comments on scientific contents ($s$) and conspiracy contents ($c$). In these settings, each user is represented by a time series $Y$ of visible events as
$$Y = \{\,\, s \quad  s \quad  s \quad  c \quad  s \quad  c \quad  s \quad  c \quad  c \quad  s \quad  c \quad c\,\,\}.$$

Notice that while the number of the visible states is fixed and determined by the available data, the number of hidden states has to be specified using prior information or common knowledge about the phenomenon under investigation. 

Formally, each HMM $\lambda$ is defined as follows:
\begin{enumerate}
	\item A set of hidden states $X = \{ \mathcal{S}, \mathcal{U}, \mathcal{C} \}$.
	\item A set of visible states (\emph{observations}) $O = \{ s,c \}$.
	\item A state transition probability matrix $\mathbf{A} = \{ a_{ij} \}$, where $a_{ij} = \mathbf{Pr}(x_{t} = X_{j} | x_{t-1} = X_{i} )$, $ i,j  \in \{\mathcal{S}, \mathcal{U}, \mathcal{C}\}$.
	\item An observation probability matrix  $\mathbf{B} = \{ b_{ki} \}$, where $b_{ki} = \mathbf{Pr}(O_{k} | X_{i} )$, $ k \in \{ s,c\}$, $i  \in \{\mathcal{S}, \mathcal{U}, \mathcal{C}\}$.
\end{enumerate}

Figure \ref{fig:fig2} provides a graphical representation of the state transition probability matrix. Notice that there is a non-zero probability of transitioning between any two states. Such a HMM is called a fully connected or ergodic HMM.

We set $\mathbf{A}$ to be an uninformative state transition probability matrix, so that $a_{ij} = 1/3$ for each $i,j \in \{\mathcal{S}, \mathcal{U}, \mathcal{C}\}$. Similarly, we set $\mathbf{B}$ to be an uninformative observation probability matrix, so that $b_{ki} = 1/2$ for each $k \in \{ s,c\}$ and $i \in \{\mathcal{S}, \mathcal{U}, \mathcal{C}\}$. Notice that one might consider to explicitly define different transition or emission probabilities according to prior knowledge.

\begin{center}
	\begin{figure}[ht]
		\centering
		\begin{tikzpicture}[->, >=stealth', auto, semithick, node distance=3cm]
		\tikzstyle{every state}=[fill=white,draw=black,thick,text=black,scale=1]
		\node[state, fill=black!10]    (S)                     {$\mathcal{S}$};
		\node[state, fill=black!10]    (U)[right of=S]   {$\mathcal{U}$};
		\node[state, fill=black!10]    (C)[right of=U]   {$\mathcal{C}$};
		\path
		(S) edge[loop left]   (S)
		edge[bend right]  (U)
		edge[bend left=60]   (C)
		
		(U) 
		edge[loop above]   (U)
		edge[bend left]  (C)
		edge[bend right]  (S)
		
		(C) edge[loop right]  (C)
		edge[bend left]  (U)
		edge[bend left=60]  (S);
		\end{tikzpicture}
		\caption{\textbf{Graphical representation of the state transition probability matrix $\mathbf{A}$.} Each node represents a hidden state, whereas each edge indicates a transition probability $p = 1/3$.}
		\label{fig:fig2}
	\end{figure}
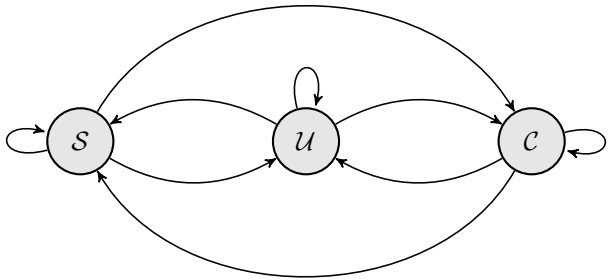
\end{center}

Each user $i \in \{1,\dots,N\}$ is represented by a temporal sequence $Y_{i}$ of comments left in response to scientific ($s$) or conspiracy ($c$) contents. For each user, we estimate the HMM parameters maximizing the likelihood of the observations sequence $\mathbf{Pr}(Y_{i}|\lambda_{i})$ by means of the Baum-Welch algorithm \cite{welch2003hidden}. In the end, each user $i$ is associated with a trained HMM $\lambda_{i}$, wherein the estimated parameters govern the transitions between hidden states and the probability to observe a sequence of comments given the hidden states of the users.

\subsection*{Constructing a similarity matrix}
Now that each user $i$ is associated to a trained HMM $\lambda_{i}$, we can construct a $N\times N$ \emph{model-based} distance matrix $\mathbf{L}$ by computing the log-likelihood value for each pair of sequences and trained HMMs via forward-backward algorithm \cite{devijver1985baum}:
$$ \mathbf{L} = \{ \ell_{ij} \}=  \{\,\,  \log\mathbf{Pr}(Y_{j} | \lambda_{i}) \,\,\}, \quad  i,j \in \{1,\dots,N\}.$$

Since such a distance matrix is not symmetric, we need to define a new symmetric distance matrix $\mathbf{D}$ by leveraging the information contained in $\mathbf{L} $:
$$ \mathbf{D} = \{ d_{ij} \} = \{\,\,  | \ell_{ii} + \ell_{jj} - \ell_{ij} - \ell_{ji} |  \,\,\}, \quad  i,j \in \{1,\dots,N\}.$$

The symmetric distance matrix $ \mathbf{D}$ represents the cross-goodness-of-fit of two sequences to the respective HMMs. Finally, we construct a similarity matrix $\mathbf{S} = \{s_{ij}\}$ by applying the following radial basis function kernel to each element of $\mathbf{D}$:

$$ s_{ij} = \begin{cases}
\exp\left( -\frac{d_{ij}}{2} \right) \qquad & i \neq j\\
0 \qquad & i = j\\
\end{cases}$$

\begin{center}
	\begin{figure}[ht]
		\centering
		\includegraphics[width = \textwidth]{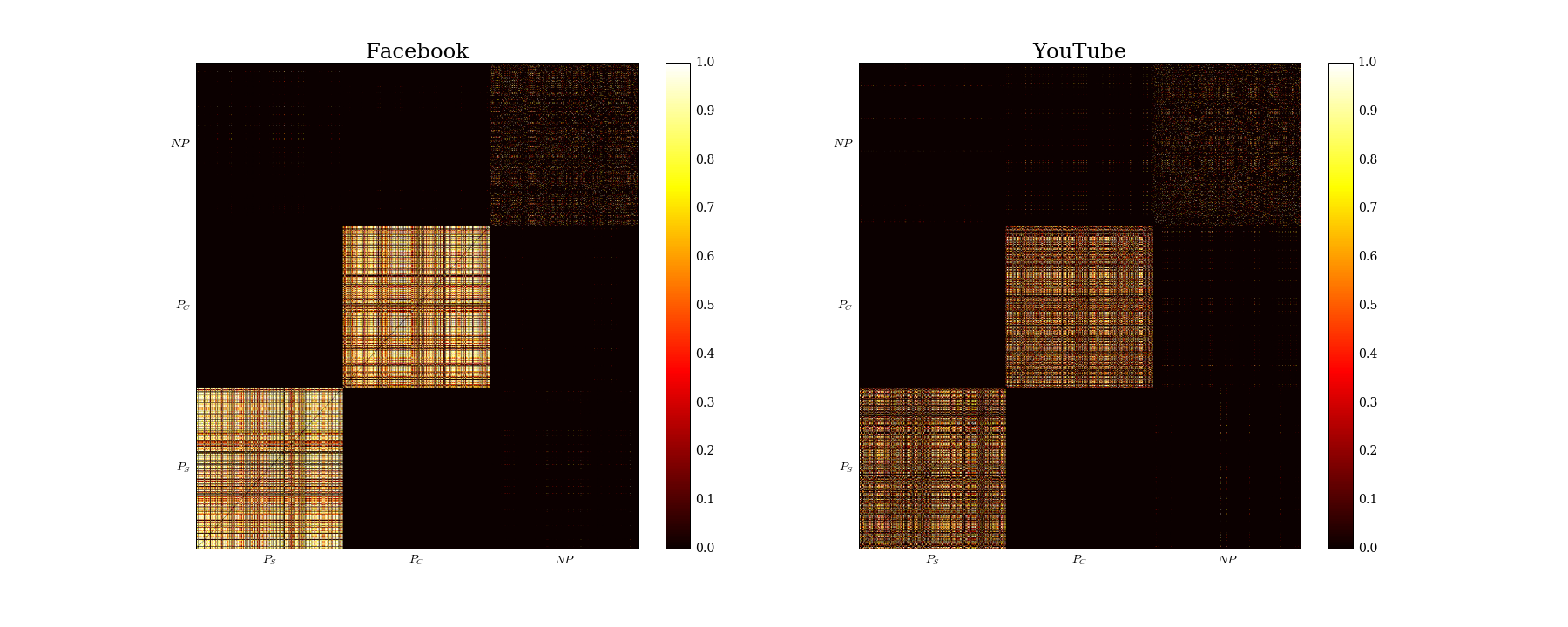}
		\caption{\textbf{Graphical representation of similarity matrices.} In both Facebook and YouTube, by means of the \emph{model-based} distance defined in the previous section, we are able to identify two large clusters of users characterized by homogeneous behavioral trajectories.}
		\label{fig:fig3}
	\end{figure}
\end{center}

Figure \ref{fig:fig3} provides a graphical representation of the similarity matrices obtained for Facebook (left panel) and YouTube (right panel) users. Such illustrations convey two interesting messages. 

First, by means of the \emph{model-based} distance defined before, we are able to identify two large clusters of users characterized by homogeneous behavioral trajectories. Such a result holds in both Facebook and YouTube, and it makes sense since the strongest similarities are observed between users supporting the same narrative --- i.e. users supporting Science ($P_{S}$) show similar behavioral trajectories, and the same apply for users supporting Conspiracy ($P_{C}$). 

Second, in both Facebook and YouTube, the similarities between not polarized users ($NP$) are weaker than the ones between polarized users ($P_{S}$ and $P_{C}$). In particular, we notice the absence of a unique homogeneous cluster of not polarized users.

To address this last observation, in the next section we apply spectral clustering methods \cite{yin2005integrating, smyth1997clustering, ghassempour2014clustering}  to the similarity matrices.

\subsection*{Applying spectral clustering}
Given a $N\times N$ similarity matrix $\mathbf{S}$, each elements $ s_{ij}$ can be viewed as the similarity between nodes $v_{i}$ and $v_{j}$. For an undirected graph $\mathcal{G}$ with nodes $v_{i}$ and edges $s_{ij}$, where $i,j \in {1,\dots,N}$, the symmetric matrix $\mathbf{S}$ is considered as the adjacency matrix for $\mathcal{G}$.

Let $k_{i} = \sum_{j \in V} s_{ij}$  be the degree of vertex $v_{i}$, and let $\mathbf{K}$ be a diagonal matrix with $k_{i}$ being its diagonal element. We can obtain a normalized stochastic matrix:
$$ \mathbf{M} = \mathbf{S}\mathbf{K}^{-1}.$$

Based on the definition of a Markov chain, $m_{ij}$ represents the transition probability of moving from $v_{i}$ to $v_{j}$. In practice, we consider a matrix
$$ \mathbf{Z} =   \mathbf{K}^{-1/2}\mathbf{M}\mathbf{K}^{1/2} =  \mathbf{K}^{-1/2}\mathbf{S}\mathbf{K}^{-1/2},$$

where $\mathbf{Z}$ is symmetric and stable in eigendecomposition. Then, the symmetric matrix $\mathbf{Z}$ can be decomposed  into the following form:
$$ \mathbf{Z} =   \mathbf{X}\mathbf{\Lambda}\mathbf{X}^{T},$$

where $ \mathbf{X}$ is a matrix obtained by stacking the eigenvectors of $\mathbf{Z}$ in columns, while $\mathbf{\Lambda} = diag(\lambda_{1},\dots,\lambda_{N})$ is a diagonal matrix with the nonnegative singular eigenvalues in descending order along the diagonal --- that is, $\lambda_{1} \geq \lambda_{2} \geq \cdots \geq \lambda_{N} \geq 0$. Since the top $E$ eigenvectors, $E \leq N$,  can capture a significant amount of information on the original data, we can map the original data into the $E$ dimensional vectors in the spectral domain, and then apply standard clustering algorithms based on the Euclidean distance such as the K-means algorithm. 

\begin{center}
	\begin{figure}[ht]
		\centering
		\includegraphics[width = \textwidth]{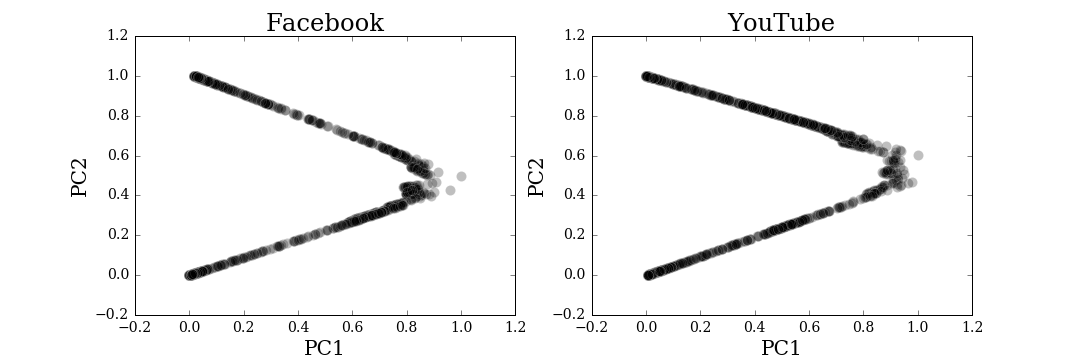}
		\caption{\textbf{Original data mapped in the spectral domain.} In both Facebook and YouTube, the eigenvectors associated with the first two eigenvalues contain enough information to let us visualize three clusters: two lines of points representing users polarized towards conflicting narratives, and a cloudy shape of points representing not polarized users.}
		\label{fig:fig4}
	\end{figure}
\end{center}

\begin{center}
	\begin{figure}[ht]
		\centering
		\includegraphics[width = \textwidth]{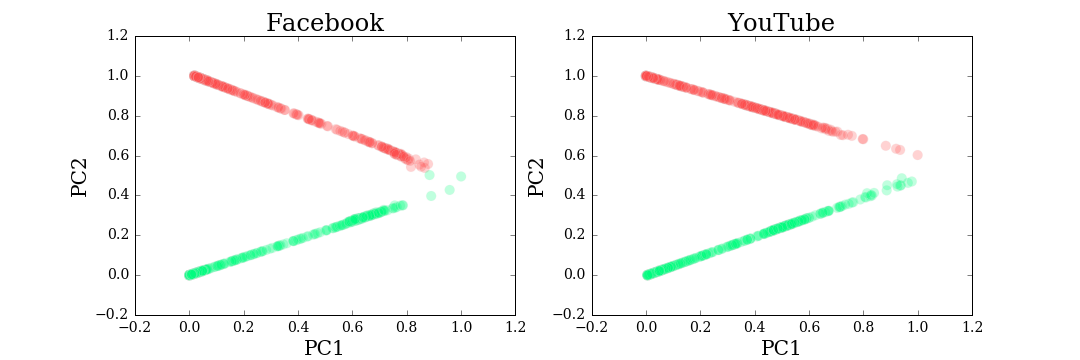}
		\caption{\textbf{K-means clustering.} In both Facebook and YouTube, K-means algorithm applied to polarized users mapped in the spectral domain clearly identifies two well separated clusters representing users polarized towards Science and users polarized towards Conspiracy.}
		\label{fig:fig5}
	\end{figure}
\end{center}

The next two figures illustrate the application of spectral clustering to the similarity matrices obtained in the previous section. Figure \ref{fig:fig4} provides a graphical representation of original data mapped in the spectral domain. The eigenvectors associated with the first two eigenvalues contain enough information to let us visualize three clusters: two almost orthogonal lines of points representing users polarized towards conflicting narratives, and a cloudy shape of points situated at the intersection of the lines, that is representing not polarized users. Figure \ref{fig:fig5} supports such an intuition by showing that the K-means algorithm applied to polarized users mapped in the spectral domain clearly identifies two well separated clusters.

\section*{Discussion}
The rise of online social media has been found responsible for distorting the collective grasp on the truth. Fake news, lies, and conspiracy theories are spreading faster than ever on Facebook, while Twitter is flooded with bots manipulating the political discussion and influencing policy preferences. Meanwhile, the natural tendency of individuals to search for information consistent with their preexisting beliefs --- \emph{confirmation bias} --- is driving the emergence of echo chambers, i.e. virtual communities wherein like-minded people reinforce their beliefs and avoid dissenting information. The concept of \emph{truth} itself is becoming more and more blurred, and someone suggests that we are entering into a \emph{post-fact} age \cite{manjoo2011true}, with obvious catastrophic consequences for democracy and society.  

In this dramatic scenario, Science has to provide a better understanding of the behavioral, cognitive, and psychological processes behind the observed dynamics, as well as develop models able to approximate and describe such processes.

In this paper, we propose a methodology that leverages Hidden Markov Models to represent behavioral trajectories of users in online social media. The intuition supporting this approach can be summarized as follows. In social network analysis, we cannot observe the actual orientations of individuals towards a specific kind of content. Indeed, the only things that we can observe are temporally ordered sequences of actions that individuals perform --- e.g. tweets, comments, likes, shares, etc. Having said that, HMMs provide a convenient as well as intuitive probabilistic framework to model the unobserved (\emph{hidden states}) orientation of individuals by taking into account their observed (\emph{visible states}) actions in online social media.

To provide platform-agnostic results, we apply our methodology to two different online social media --- Facebook and YouTube --- showing that our approach is able to discover homogeneous clusters of individuals showing similar behavioral trajectories in both the platforms considered. 

Clearly, we have to point out some limitations of the present work. First, given the complexity of human behavior, any attempt to infer the actual orientation of individuals as well as their motivations is out of the scope of the proposed methodology. Nevertheless, we think that our approach yields a useful approximation and representation of behavioral trajectories that can be exploited to identify homogeneous clusters of users. Then, we have to emphasize that our dataset is a particular dataset, and thus we cannot venture any general claims. Context matters, and far more research would be necessary to support any such general claims. A further limitation of the present study is the computational time required to train HMMs for a large number of individuals. Still, we believe that the proposed methodology is straightforward to implement, and supported by sound and intuitive theoretical foundations. Moreover, HMMs provide a flexible probabilistic framework that can adapt to different contexts.

In particular, we think that our methodology could be used to investigate the effectiveness of fact-checking \cite{ciampaglia2015computational,tambuscio2016network} and debunking of false information proliferating in online social media. Some studies pointed out the inefficacy of debunking and the concrete risk of a backfire effect from the most committed partisans \cite{nyhan2010corrections, zollo2015debunking,redlawsk2002hot,gollust2009polarizing,nyhan2013hazards,schaffnermisinformation}. Differently, recent studies found that individuals heed factual information, even when such information challenges their partisan and ideological attachments \cite{wood2016elusive,weeks2015emotions,nyhannda}. In this scenario, we think that our approach might help in the identification of homogeneous clusters of individuals showing similar reactions and behaviors with respect to debunking and fact-checking. 

\section*{References}

\bibliographystyle{unsrt}
\bibliography{biblio}

\end{document}